# Direct observation of a photoinduced topological phase transition in Bi-doped (Pb,Sn)Se

Masataka Mogi,[1,§,†] Dongsung Choi,[2,†] Louis Primeau,[3] Baiqing Lv,[1,4] Doron Azoury,[1] Yifan Su,[1] Liang Fu,[1] Yang Zhang,[3,5] Nuh Gedik[1,*]

[1]Department of Physics, Massachusetts Institute of Technology, Cambridge, MA, USA.

[2]Department of Electrical Engineering and Computer Science, Massachusetts Institute of Technology, Cambridge, MA, USA.

[3]Department of Physics and Astronomy, University of Tennessee, Knoxville, TN, USA.

[4]School of Physics and Astronomy, Shanghai Jiao Tong University, Shanghai, China.

[5]Min H. Kao Department of Electrical Engineering and Computer Science, University of Tennessee, Knoxville, TN, USA.

[§]Present address: Department of Applied Physics, University of Tokyo, Bunkyo-ku, Tokyo, Japan.

[†]These authors contributed equally to this work: Masataka Mogi, Dongsung Choi.

[*]e-mail: gedik@mit.edu

# Abstract

Ultrafast photoexcitation offers a novel approach to manipulating quantum materials. One of the long-standing goals in this field is to achieve optical control over topological properties. However, the impact on their electronic structures, which host gapless surface states, has yet to be directly observed. Here, using time- and angle-resolved photoemission spectroscopy, we visualize the photo-induced evolution of the band structure in $Bi_y(Pb_{1-x}Sn_x)_{1-y}Se(111)$ films from topological to trivial insulators. Following near-infrared ultrafast laser excitation, we observe that the topological surface state opens a substantial gap of up to 0.1 eV. Considering the topological phase diagram associated with lattice distortion and atomic displacement, we show that a uniaxial strain generated by the ultrafast optical pulse is sufficiently effective and strong for the observed topological phase transition. Our study highlights the potential of optical tuning of materials through laser excitation to control topological properties on ultrafast timescales.

# Main text

The exploration and manipulation of topological phases of matter have emerged as one of the most exciting frontiers in condensed matter physics, motivated by their unique charge and spin transport properties [1,2]. Various static approaches, including modification of composition [3-5], application of pressure [6,7], strain [8], and electric fields [9,10], have successfully adjusted material parameters such as spin-orbit coupling, lattice constant, and energy level shift, which triggers bulk band inversion and topological phase transitions. In particular, transitions between topological insulators (TIs) and topologically trivial insulators are important due to the prospect of switching the topological surface states on and off, potentially enabling the development of next-generation memory and transistor devices utilizing topological properties.

Recently, ultrafast photoexcitation has attracted considerable attention as a novel method for dynamically manipulating topological phases. A prime example is the use of Floquet-Bloch states, driven by coherent light-matter interaction [11,12]; for instance, circularly polarized light can induce a gap in the surface state of TIs by breaking time-reversal symmetry [12]. However, Floquet-Bloch states can only exist while the light pulse is active, posing a challenge to the stabilization of the photo-induced phases. Another intensively explored approach is to modify the crystal structure of materials. Previous studies [13-16] have reported that crystal symmetry can be altered through optical pulse-induced shear strains of the lattice, potentially facilitating the transition from 3D topological Weyl semimetals to trivial metals. Nevertheless, existing evidence for topological phase transitions has primarily relied on electronic structure calculations. The small energy (< 50 meV) and momentum scales (< 0.05 Å$^{-1}$) of the Weyl point motion embedded in bulk electronic states make its direct observation using angle-resolved photoemission spectroscopy (ARPES) challenging [13,17,18].

Here, we report the direct observation of a transition from a TI to a trivial insulator in the Bi$_y$(Pb$_{1-x}$Sn$_x$)$_{1-y}$Se system by exciting with ultrashort, infrared light pulses (135 fs pulse duration, energy of 1.2 eV). A band gap emerges in the trivial insulator state, as large as 130 meV, which is markedly larger than the gap previously realized by Floquet engineering in Bi$_2$Se$_3$ (~ 50 meV) [12]. Pb$_{1-x}$Sn$_x$Se is a prototypical topological crystalline insulator (TCI) [5,6,9,19,20], featuring four 2D Dirac cones in the surface Brillouin zone [Fig. 1(a,c)]. The crystal symmetry of the face-centered cubic structure characterizes the topological invariant, known as the mirror Chern numbers, which can be manipulated by controlling the composition ($x$), temperature [5,21], electric field [9,20,22], pressure [6,7], and strain [23,24]. Specifically, Bi doping induces a topological phase transition from TCI [Fig. 1(c)] to a time-reversal-invariant $\mathbb{Z}_2$ TI phase [Fig. 1(d)], where one of the surface Dirac cones at the $\bar{\Gamma}$ point becomes gapped due to rhombohedral distortion along the [111] direction [25]. Owing to its simple band structure near the Fermi energy ($E_F$), the Bi$_y$(Pb$_{1-x}$Sn$_x$)$_{1-y}$Se system serves as an excellent platform for exploring novel photo-induced changes in band topology [26].

We grew the Bi$_y$(Pb$_{1-x}$Sn$_x$)$_{1-y}$Se thin-film samples with (111)-oriented surfaces on InP(111) substrates [27]. We focus on the samples with the composition $x$ = 0.21 to 0.24, $y$ = 0 to 0.05. Within this range, a topological phase transition is expected to be driven by temperature ($T$) at around $T \sim 100$ K [5,21,25]. We use extreme ultraviolet femtosecond light source-based time-resolved ARPES to access the entire Brillouin zone [37], providing evidence of a photo-induced transition from $\mathbb{Z}_2$ TI [Fig. 1(d)] to a trivial insulator [Fig. 1(b)].

We first characterize the electronic structure of our samples. In an undoped $Pb_{1-x}Sn_xSe$ ($x = 0.23$, $y = 0$) film, the band at the $\bar{\Gamma}$ point exhibits a gap at room temperature [Fig. 1(e)], but it becomes gapless at low temperatures ($T = 45$ K) [Fig. 1(f)], signifying the trivial phase to TCI phase transition. In contrast, Bi doping ($y \sim 0.05$) results in the gap opening as large as 120 meV at the $\bar{\Gamma}$ point, even at low temperature ($T = 35$ K) [Fig. 1(g)]. The W-shaped conduction band, which appears when Bi is doped, indicates Rashba splitting due to surface and/or bulk inversion symmetry breaking [22]. At the $\bar{M}$ point, on the other hand, the gap closes at low temperature ($T = 50$ K), showing a Dirac-like dispersion [Fig. 1(h)], similar to what we observe in the undoped (Pb,Sn)Se. Note that because the Fermi level ($E_F$) is positioned close to the Dirac point (DP), we conducted these measurements using pump excitation at time zero ($\Delta t = 0$ ps) – the point of temporal overlap between pump and probe pulses [27] – to observe the upper band by populating the unoccupied states while minimizing the effect of lattice heating. Our results are consistent with the previous study [25] that identified the Bi-doped sample as being the $\mathbb{Z}_2$ TI phase [Fig. 1(d)].

To elucidate the relationship between the observed electronic structure and the crystal structure of the Bi-doped sample [Fig. 1(i)], we conducted density functional theory (DFT) band structure calculations [Fig. 1(j)]. Our calculations take into account both the rhombohedral distortion, characterized by the rhombohedral angle α (with α = 60° signifying a face-centered cubic structure), and atomic displacement, denoted by the distance fraction δ [Fig. 1(a)], which represents inversion symmetry breaking. Previous theoretical work [40] has shown that the TCI phase possesses gapless surface states at the $\bar{\Gamma}$ and $\bar{M}$ points [Fig. 1(c)], while the $Z_2$ TI phase only possesses these states at the $\bar{M}$ points [Fig. 1(d)] in the presence of rhombohedral distortion. These $\bar{\Gamma}$ and $\bar{M}$ points correspond to Z and L points in the bulk Brillouin zone of the rhombohedral lattice [Fig. 1(a)]. While the previous study further presented the topological phase diagram as a function of ferroelectric distortion, coupling α and δ, our DFT calculations independently map out the results in Fig. 1(k) as a function of α and δ by calculating the bulk band gap size at the Z and L points [Fig. 1(j)] as well as the mirror Chern number [27]. From these calculations, we newly identify two key features: First, undoped (Pb,Sn)Se, with inversion symmetry (δ = 0), robustly maintains the TCI phase even with minor rhombohedral distortions; second, achieving the $\mathbb{Z}_2$ TI phase requires rhombohedral distortion or atomic displacement. Considering that the distortion angle, characterized by α – 60°, in our Bi-doped (Pb,Sn)Se films is as low as 0.15°, as measured by x-ray diffraction [27], inversion symmetry breaking emerges as a critical factor for the realization of the $\mathbb{Z}_2$ TI phase in the Bi-doped (Pb,Sn)Se.

Having confirmed the $\mathbb{Z}_2$ TI electronic structure in our Bi-doped (Pb,Sn)Se film and evaluated the impact of Bi doping on the lattice, we now photoexcite the Bi-doped (Pb,Sn)Se film using ultrashort, intense near-infrared pulses (~3 mJ cm$^{-2}$, 1.2 eV). We performed time-resolved ARPES measurements with a delay time, $\Delta t$, for each pump pulse at a repetition rate of 300 kHz. Figure 2(a) displays the ARPES spectra at selected $\Delta t$ around the $\bar{\text{M}}$ point. Remarkably, we observe a gap opening at positive delay times ($\Delta t > 2$ ps, and possibly even at $\Delta t = 0.2$ ps [27]), which persists at $\Delta t = 6$ ps, whereas the band is gapless at $\Delta t = 0$ ps. Given that the band at $\bar{\Gamma}$ is initially gapped [Fig. 1(g)] and persists after photoexcitation as shown in Fig. S5(a) [27], this photo-induced gap-opening suggests a topological phase transition from $\mathbb{Z}_2$ TI [Fig. 1(d)] to a trivial insulator [Fig. 1(b)].

The gap-opening feature becomes evident when we extract energy distribution curves (EDCs) [Fig. 2(c)] at the $\bar{\text{M}}$ point. An intensity dip near $E_F$ appears at positive delay times, with the energy scale estimated to be around 130 meV, corresponding to the formation of an energy gap. Although we observe finite intensity in the dip, which is especially significant at 2 ps, we attribute it to photo-excited in-gap states, a common observation in doped semiconductors, such as doped topological insulators [38,39]. The appearance of the dip structures resembles the effect of lattice heating induced by the laser. As shown in the inset of Fig. 2(c), intensity dips appear at 150 K and 300 K, measured at Δ*t* = 0 ps. Since the lattice heating effect should not be exclusive to the Bi-doped case, we also measured the undoped samples; however, we did not observe the gap-opening behavior at either $\bar{\mathrm{M}}$ [Fig. 2(b)] or $\bar{\Gamma}$ [Fig. S5(a)] [27] despite using the same fluence (3 mJ cm$^{-2}$) and higher absorbance (see Figs. S3 and S6 [27]). All the EDCs shown in Fig. 2(d), with varying delay times, resemble the time-zero data of the Bi-doped sample [Fig. 2(c)], indicating that the observed photo-induced band modification is not merely due to an increase in the sample temperature.

Next, we examine the relaxation dynamics of the photo-excited Bi-doped and undoped samples. Figure 3(a) shows the time evolution of spectral intensity of the Bi-doped sample at the $\bar{\mathrm{M}}$ point. Following photoexcitation, the spectral weight near the Dirac point (DP) quickly decreases, but the spectral weight above the DP persists [Fig. 3(b)]. We extract the decay time using a single exponential decay model described by $f(t) = I_{peak}\exp\left(-\frac{t-t_0}{\tau}\right)\left(1 + \mathrm{erf}\left(\frac{2\sqrt{2}(t-t_0)}{w}\right)\right) + C$, where $I_{peak}$, $t_0$, erf, $w$, $C$, and $\tau$ are the peak intensity, time zero, error function, response time, constant, and decay time, respectively. As shown in Fig. 3(c), the decay time becomes longer (up to 10 ps) as the energy decreases, but near the Dirac point, the decay time is as short as 2 ps. This suggests that the nature of the carriers differs between those at the conduction band and those near the Dirac point, indicating that the gap is open and that the photo-excited in-gap states have a short lifetime.

In contrast, the intensity of the undoped sample decays much faster, taking about 2 ps for the electrons excited above the Dirac point, as shown in Fig. 3(d,e,f). This contrast reflects the presence or absence of a gapless state and the persistence of the photo-induced trivial insulator phase. In general, when the gap size is large compared to the phonon energy scale (< 20 meV) [41], the photo-excited electrons in the conduction band are unable to transit back to the valence band via electron-phonon scattering channels. Based on these results, the observed band dynamics upon photoexcitation can be interpreted as follows: Near time-zero (Δ*t* < 0.2 ps; see Fig. S12 [27]), the electrons are excited while the lattice remains unheated. Therefore, our ARPES spectra at this stage reflect the band structure with minimal modification from the equilibrium state. Once the electron-phonon interaction is activated, it can alter the lattice structure and consequently modify the band structure. After the lattice cools down - a timescale for heat dissipation that is limited by sound propagation in the lattice (typically nanoseconds) - we would observe the band structure with a gapless state, with a timescale of at least less than 3 μs according to our laser repetition rate of 300 kHz.

We now turn to discussing the mechanism behind the photo-induced topological phase transition observed exclusively in the Bi-doped sample. Based on our DFT calculations [Fig. 1(k)], one conceivable explanation is the enhancement of atomic displacement (δ), which could lead to a transition from $\mathbb{Z}_2$ TI to a trivial insulator. Various pathways can modulate atomic displacement by photoexcitation, such as the generation of an $A_{1g}$ Raman mode [42,43] and the surface space-charge effect [44], as previously reported in ferroelectric GeTe. In the case of the $A_{1g}$ Raman mode, cation and anion atoms move oppositely along the [111] direction, modulating ferroelectricity [45]. This atomic displacement, which breaks inversion symmetry, could be detected through the modulation of Rashba splitting via spin-orbit interaction. However, in our experiment, the Rashba-like band splitting observed at the $\bar{\Gamma}$ point [27] remains unchanged upon photoexcitation. Thus, an enhancement of inversion symmetry breaking through this scenario is not enough to explain the observed transition. Other potential mechanisms for controlling the band gap could be strain and lattice vibrations via electron-phonon coupling. Studies on the origin of equilibrium temperature-dependent gap size changes [46,47] have demonstrated that thermal lattice expansion accounts for about half of the gap size change, while the impact of lattice vibrations is comparably significant. In the following paragraph, we demonstrate that even without considering the impact of lattice vibrations promoted by ultrafast photoexcitation, which is challenging to quantify, photoexcitation-driven lattice strain is sufficiently effective to control the topological phase.

Upon photoexcitation, the lattice experiences instantaneous stress through energy transfer between electrons and phonons. However, the lateral strain of the lattice is restricted to the timescale necessary for sound propagation to exceed our pump-laser footprint (~100 μm) [49]. Consequently, the lattice strain primarily occurs out-of-plane, that is, along the [111] direction, and the lattice constant changes uniaxially. Such ultrafast out-of-plane lattice strain, occurring within a few picoseconds, has been observed in ultrafast x-ray and electron diffraction [48,49,50]. The strain level is estimated to reach several percent of expansion in our system [27], where ultrafast changes in electronic energy are more effective at generating strain than simply increasing the lattice temperature (~0.1%). We propose that this anisotropic deformation of the lattice may result in different photo-induced dynamics between Bi-doped and undoped (Pb,Sn)Se in terms of band topology, due to the presence of atomic displacement along the [111] direction.

To corroborate our hypothesis, we calculated the 3D bulk electronic band structures and their band topology under variations of uniaxial lattice expansion along [111] ($c$) as well as uniform lattice expansion ($\lambda$). The number of gapless surface states in the surface Brillouin zone changes when the bulk energy gap closes at the Z or L points in the 3D Brillouin zone [Fig. 1(b)], signalling a topological phase transition. We start with the lattice constant where the TCI phase is obtained ($\lambda = \lambda_0$ or $c = c_0$). In the case of isotropic lattice expansion, intended to simulate a temperature increase, TCI ($\delta = 0\%$) [Fig. 4(a)] and $\mathbb{Z}_2$ TI ($\delta = 1.5\%$) [Fig. 4(b)] to trivial insulator transitions occur at $\lambda/\lambda_0 = 1.03$ and 1.005, respectively. The difference in the transition values is because, with a fixed lattice parameter, a finite $\delta$ brings the system closer to the trivial insulator, as shown in Fig. 1(k). On the other hand, as shown in Fig. 4(d), uniaxial lattice expansion results in a similar phase transition only when the inversion symmetry is broken. However, when the inversion symmetry is maintained, this transition cannot be achieved up to $c/c_0 = 1.06$ expansion [Fig. 4(c)], which means that the TCI phase is stable against ultrafast uniaxial deformation. Therefore, we conclude that the presence of inversion symmetry breaking facilitates the observed photo-induced topological phase transition.

To summarize, our results demonstrate a photo-induced topological phase transition in Bi-doped (Pb,Sn)Se films by directly visualizing the band structure. The ultrafast photoexcitation plays an important role in inducing uniaxial (out-of-plane) strain in the lattice, an effect that cannot be achieved by simply increasing the temperature which uniformly expands the lattice. The resultant topological phase transition has been achieved by Bi-doping, breaking inversion symmetry, which is an additional capability to control materials parameters in (Pb,Sn)Se. Integration of such extensive material parameter tuning and tailoring ultrafast laser excitation could open a way for the development of ultrafast manipulation of topological properties and potential optoelectronic applications.

We thank Nikesh Koirala and Bryan Fichera for experimental support and Martin Eckstein, Edoardo Baldini, Honglie Ning, Haoyu Xia and Yukako Fujishiro for fruitful discussions. This work was supported by the US Department of Energy, BES DMSE (data taking, analysis and manuscript writing) and Gordon and Betty Moore Foundation's EPiQS Initiative grant GBMF9459 (instrumentation). M.M. acknowledges support from JSPS Overseas Research Fellowships and JST PRESTO (no. JPMJPR23HA). Y.Z. is supported by the start-up fund at University of Tennessee Knoxville.

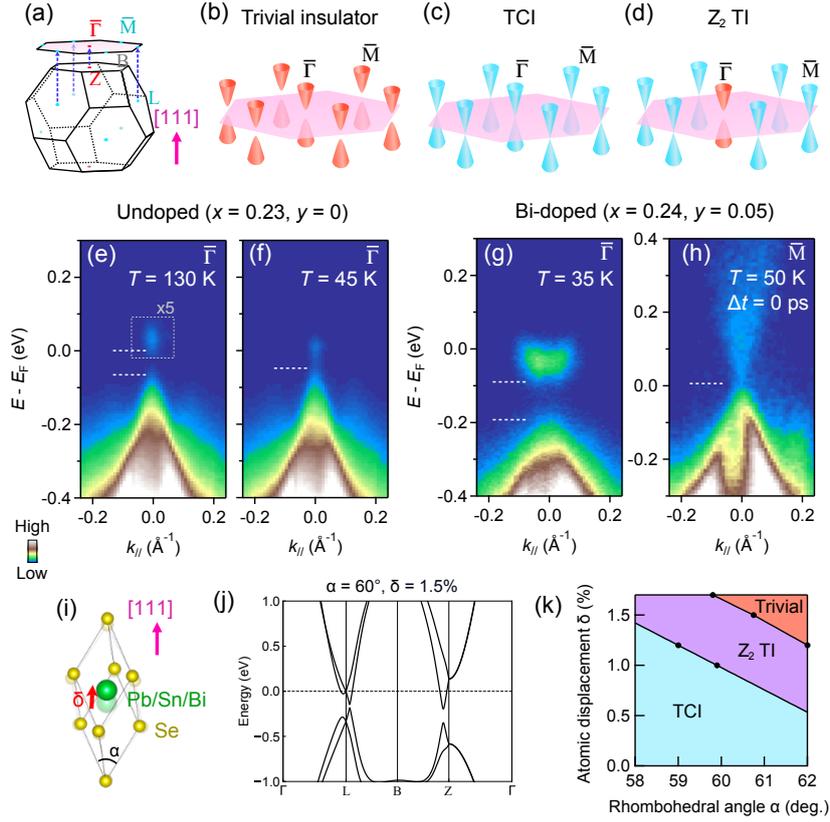

FIG. 1. (a) Brillouin zone of (Pb,Sn)Se with rhombohedral distortion along [111] and the projected (111) surface Brillouin zone. (b-d) Band diagrams of the trivial insulator phase (b), topological crystalline insulator (TCI) phase (c), and $\mathbb{Z}_2$ topological insulator (TI) phase (d). (e-h) ARPES spectra for undoped $(Pb_{0.77}Sn_{0.23})Se$ around the $\bar{\Gamma}$ point, measured at $T = 130$ K (e) and at $T = 45$ K, and for Bi-doped $(Pb_{0.76}Sn_{0.24})Se$ measured at the $\bar{\Gamma}$ (g) and $\bar{M}$ (h) points. The white broken lines are guides to the eye, indicating whether the band is gapless or gapped. (i) Schematic of rhombohedral distortion along the [111] direction in (Bi,Pb,Sn)Se crystal. The distortion is characterized by the rhombohedral angle ($\alpha$) and atomic displacement ($\delta$). (j) DFT-calculated 3D bulk band structure for $(Pb_{0.77}Sn_{0.23})Se$ with $\alpha = 60°$ and $\delta = 1.5\%$. (k) Topological phase diagram for rhombohedral angle $\alpha$ and atomic displacement $\delta$, derived from DFT calculations. Each topological phase is determined by the Wilson-loop calculation [27].

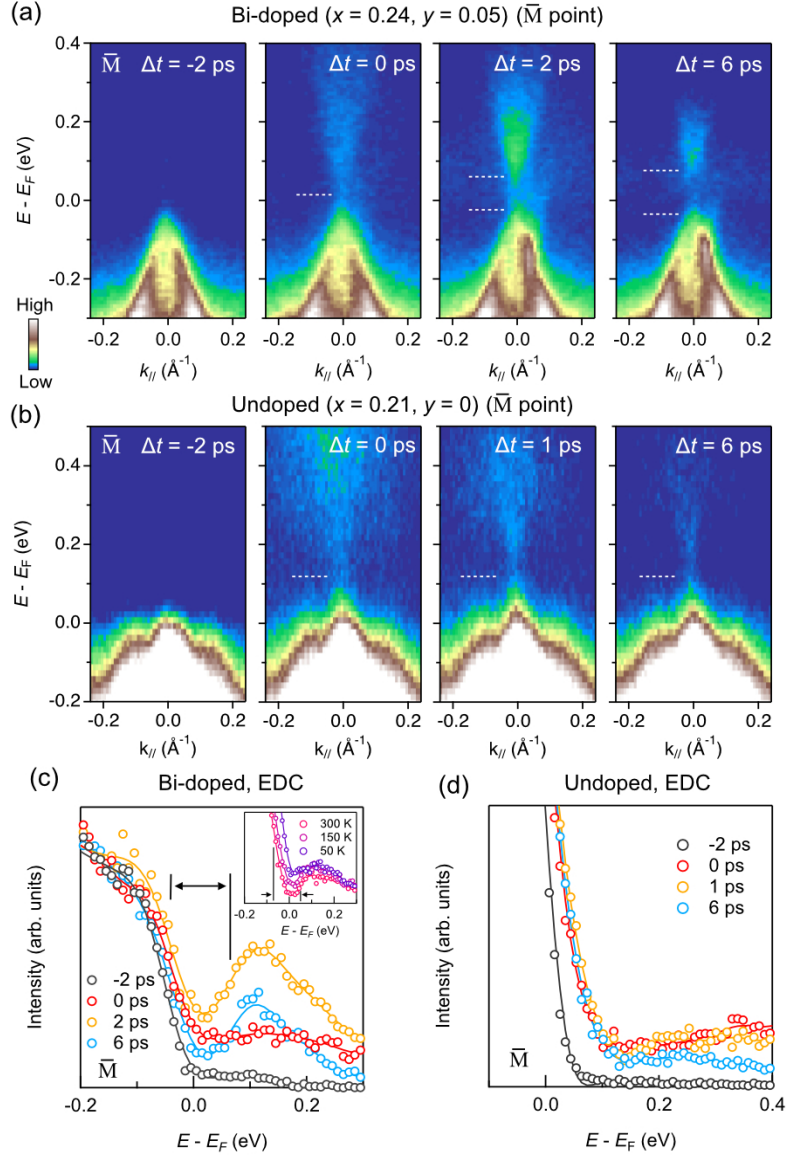

FIG. 2. (a,b) Time-resolved ARPES spectra measured around the $\bar{M}$ point with various pump-probe delay times $\Delta t$ in Bi-doped $(Pb_{0.76}Sn_{0.24})Se$ (a) and undoped $(Pb_{0.79}Sn_{0.21})Se$ (b) thin films. (c,d) Energy distribution curves (EDCs) at the $\bar{M}$ point taken from (a,b), respectively. The inset shows EDCs at $\Delta t = 0$ ps for various temperatures ($T = 50$, 150, and 300 K). The gap size is estimated by the distance between the inflection points of EDCs. The incident pump fluence at the sample surface is ~3 mJ cm$^{-2}$.

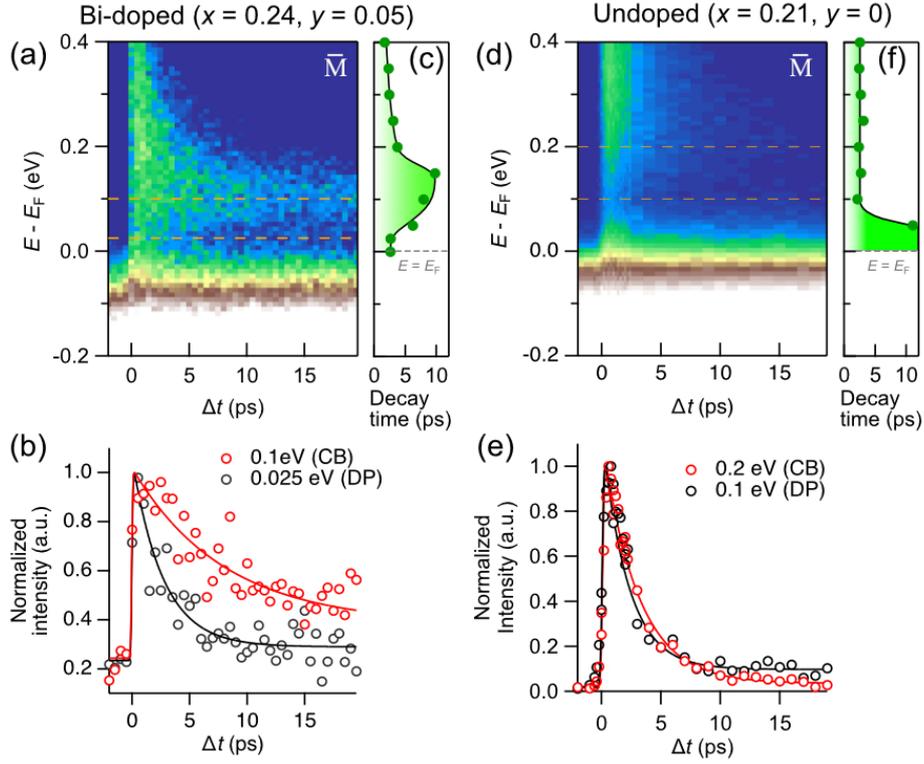

FIG. 3. (a,d) Time-evolution of the spectral intensity for the $\bar{M}$ point in the Bi-doped (a) and undoped (d) samples. (b,e) Intensity profile as a function of $\Delta t$ at the Dirac point (DP) and at the conduction band (CB) region, ~ 0.1 eV above the DP for the Bi-doped (b) and undoped (e) samples. Solid curves are fits to a single-exponential decay model (see the main text). (c,f) Energy ($E - E_F$) dependence of the decay time extracted from the fittings. Solid curves are guides to the eye.

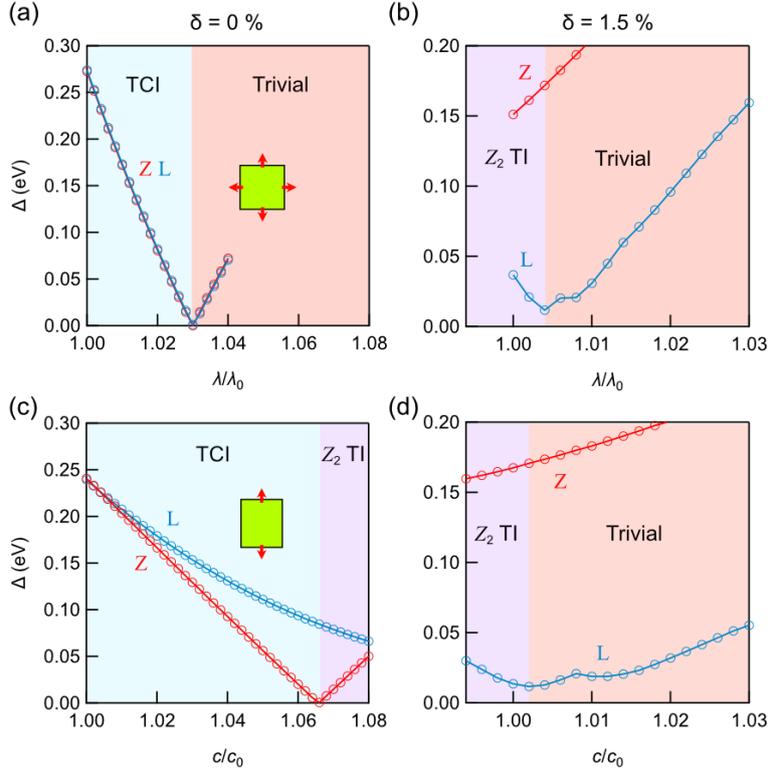

FIG. 4. (a,b) Isotropic lattice expansion ratio ($\lambda/\lambda_0$) dependence of the energy gap size ($\Delta$) of the 3D bulk bands for (Pb,Sn)Se without (a) and with (b) atomic displacement ($\delta = 1.5\%$) at the Z and L points. (c,d) Uniaxial lattice expansion ratio ($c/c_0$) dependence of $\Delta$ for (Pb,Sn)Se without (c) and with (d) atomic displacement ($\delta = 1.5\%$) at the Z and L points. Note that the gap at the L point for (b) and (d) does not reach zero near the topological phase boundary because a Weyl semimetal phase appears owing to inversion symmetry breaking, in which the zero gap states appear at momenta slightly away from the L point, although our ARPES spectra do not have the momentum resolution to observe it. Each topological phase is determined by the Wilson-loop calculation [27].

17